# Ultra-broadband suppression of sound scattering via illusion metamaterials

Chenkai Liu[1,2,†], Chu Ma[3,†], Yun Lai[1,*] and Nicholas X. Fang[2,4,*]

[1] MOE Key Laboratory of Modern Acoustics, National Laboratory of Solid State Microstructures, School of Physics, and Collaborative Innovation Center of Advanced Microstructures, Nanjing University, Nanjing 210093, China

[2] Department of Mechanical Engineering, Massachusetts Institute of Technology, 77 Massachusetts Avenue, Cambridge, MA 02139, USA

[3] Department of Electrical and Computer Engineering, University of Wisconsin-Madison, WI 53706, USA

[4] Department of Mechanical Engineering, University of Hong Kong, Pokfulam Road, Hong Kong

[†] C. L. and C. M. contributed equally to this work.

[*] Corresponding author. Emails: laiyun@nju.edu.cn (Y. L.), nicxfang@hku.hk (N. X. F.)

The scattering of waves is a ubiquitous phenomenon in physics, yet there are numerous scenarios, such as the pursuit of invisibility, where suppressing it is of utmost importance. In comparison to prior methods which are restricted by limited bandwidths, here we present a technique to suppress sound scattering across an ultra-broad spectrum by utilizing illusion metamaterials. This illusion metamaterial, consisting of subwavelength tunnels with precisely crafted internal structures, has the ability to guide acoustic waves around the obstacles and recreate the incoming wavefront on the exit surface. Consequently, two ultra-broadband illusionary effects are produced: "disappearing space" and "time shift". Simultaneously, all signs of sound scattering are removed across an exceptionally wide spectrum, ranging from the quasistatic limit to an upper limit of the spectrum, as confirmed by full-wave simulations and acoustic experiments. Our approach represents a major step forward in the development of broadband functional metamaterials and holds the potential to revolutionize various fields, including acoustic camouflage and reverberation control.

## INTRODUCTION

Wave scattering, as illustrated in Fig. 1a, is a ubiquitous wave phenomenon that can cause significant effects, such as energy flux redistribution, diffraction (*1*, *2*) and diffusion (*3*), etc. Despite its impact, many real-world applications require techniques to eliminate or minimize the consequences of wave scattering. For example, one well-known example is the invisibility cloak (*4*, *5*), which was developed with the advancement of artificial metamaterials (*6*–*12*). By manipulating acoustic waves with the extraordinary acoustic metamaterials (*13*–*17*) and metasurfaces (*18*), acoustic invisibility cloaks (*19*–*21*) have been demonstrated. From the perspective of illusionary effects of metamaterials (*22*–*26*), transmission-type (*27*–*31*) and reflection-type (*32*–*34*) cloaks both realize the special illusion of "empty space", without or with a reflecting plane, respectively. Additionally, there are several other methods for scattering suppression (SS), such as zero-index media (*35*–*37*), interference (*38*), topological effects (*39*), and spatial dispersion (*40*), etc. Nevertheless, it's worth noting that these approaches may have a limited narrow operating bandwidth due to the causality principle and resonant nature of structures, which could potentially hinder their practical applications.

In this work, we overcome the longstanding bandwidth limitation in sound scattering suppression by creating a class of illusion metamaterials that exhibit two broadband illusionary effects: "*disappearing space*" and "*time shift*", instead of the "empty space" in invisibility cloaks and other structures. Such illusion metamaterials, consisting of subwavelength acoustic tunnels that route waves around the obstacles, can work as SS devices with unprecedented broad bandwidths. Attributed to the equal acoustic path design of each tunnel and the uniform geometric shape of the entry and exit surfaces, the wavefront of incident wave is accurately reproduced on the exit surface over an ultra-broad spectrum. Consequently, an illusion of the space occupied by the metamaterial and obstacles have completely "disappeared", while at the same time, the time is "shifted" backward by a certain amount. Through full-wave simulations and acoustic experiments, we demonstrate that this ultra-broadband SS device works across all frequencies from the quasistatic limit to an upper limit. In one example, we show that the bandwidth covers the entire regime from 1 kHz to 16 kHz. Our work replaces the traditional concept of SS as creating "empty space" by a new paradigm of "*disappearing space*" and "*time shift*", enabling the long-coveted suppression of scattering with the extreme bandwidth and frequency-independent feature, which was previously impossible.

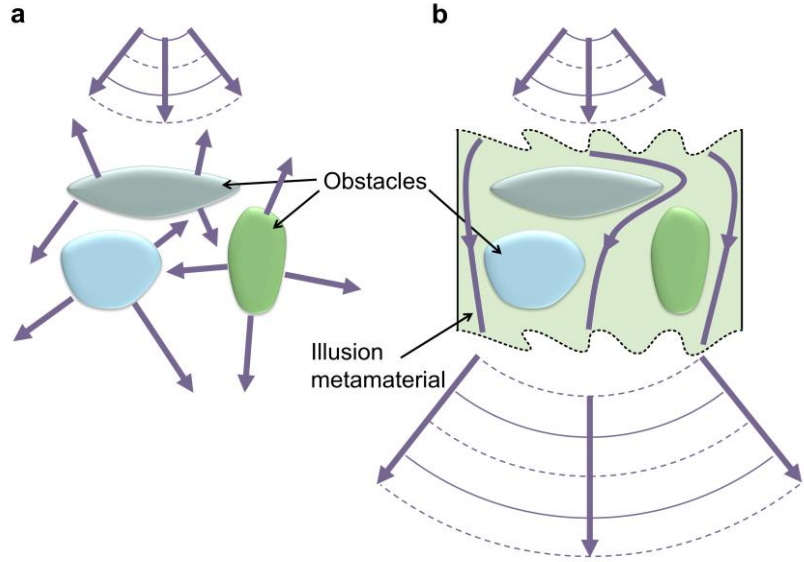

**Fig. 1. Illustration of the illusion metamaterial as a SS device.** Upon the incidence of a sound wave, **a** multiple scattering is almost always generated by a random arrangement of obstacles, however, **b** when the obstacles are embedded in the illusion metamaterial (SS device), the backward reflection is removed and the waves can propagate around the obstacles with their wavefronts undistorted. The arrows and the alternating solid and dashed lines represent sound waves and their wavefronts, respectively.

## RESULTS

**Acoustic tunnels with internal protrusion structures**

To show the generality of our approach, the tunnel is designed to possess an arbitrary shape, as shown in Fig. 2a. Suppose the tunnel has a varying cross-section-width $w$, and a horizontal length $a$. To achieve a tunable acoustic path, protrusion structures with a height of $h$ and a lattice constant of $d$ are added to the interior of the tunnel. The thickness of the tunnel shell and the protrusion is set as $t$. When the width of the tunnel is much smaller than the wavelength, there is only one mode in the tunnel and the acoustic tunnel can be described by a transmission-line model. A collection of such tunnels leads to anisotropic metamaterials (*41–43*) that guide sound along certain directions. Different from the simple tunnels utilized in the previous works, here the internal protrusion structures play a key role in modulating the total acoustic path of each tunnel.

To illustrate this mechanism, we have performed full-wave simulations for the tunnel plotted in Figs. 2b-c. The geometric parameters are set as $t = 1$ mm, $d = 5$ mm and $a = 20$ cm. The maximum and minimum width of the tunnel are set as $8.95$ mm and $3.95$ mm, respectively. A frequency of $f = 6860$ Hz is chosen for demonstration. Figure 2b shows the acoustic pressure field distributions inside the tunnels with different geometric parameters of $h/w$. It is clearly seen that the phase shift of the transmitted wave increases substantially when $h/w$ increases. We note that $w$ and $h$ simultaneously change along the propagation path, but the ratio of $h/w$ is fixed in one tunnel. Figure 2c presents the phase shift $\Delta\phi$ and the transmittance $|t|^2$ as a function of $h/w$ at $f = 6860$ Hz. We can observe that the transmitted phase shift can be conveniently modulated to cover the range of $2\pi$ by changing $h/w$ from 0 to 0.32, while the average transmittance is kept above $97\%$, indicating that such type of tunnels can achieve a tunable acoustic path with a high efficiency. Investigation on some minor factors to the transmittance, including the incident angle of sound, the bending angle and the variation width of the tunnel, are discussed in the Supplemental Materials. The results demonstrate that the functionality of the tunnel is quite robust against these factors. We note that comparing with other designs based on phase manipulation, such as the zigzag structures (*35*), the protrusion structures exhibits a better performance in impedance matching.

Interestingly, such a type of tunnels with a curved path can be approximated as straight tunnels with a slower and dispersion-free effective sound speed and this works in an ultra-broad spectrum. The transmission phase of the tunnel is described as $\phi = k_0 L$, where $k_0$ is the wave number in the background (air) and $L = n_r a$ is the acoustic path. Here, $n_r = \phi/(k_0 a)$ is the effective refractive index. In Fig. 2d, we plot the calculated $n_r$ as a function of $h/w$ and the frequency $f$ for the tunnel in Fig. 2a. It is clearly seen that $n_r$ increases with the increase of $h/w$, and is almost unchanged from 0.1 kHz to 15 kHz. And a larger $n_r$ exhibits a stronger dispersion in the frequency region. We emphasize that the frequency-independent refractive index is the key to the realization of ultra-broadband SS.

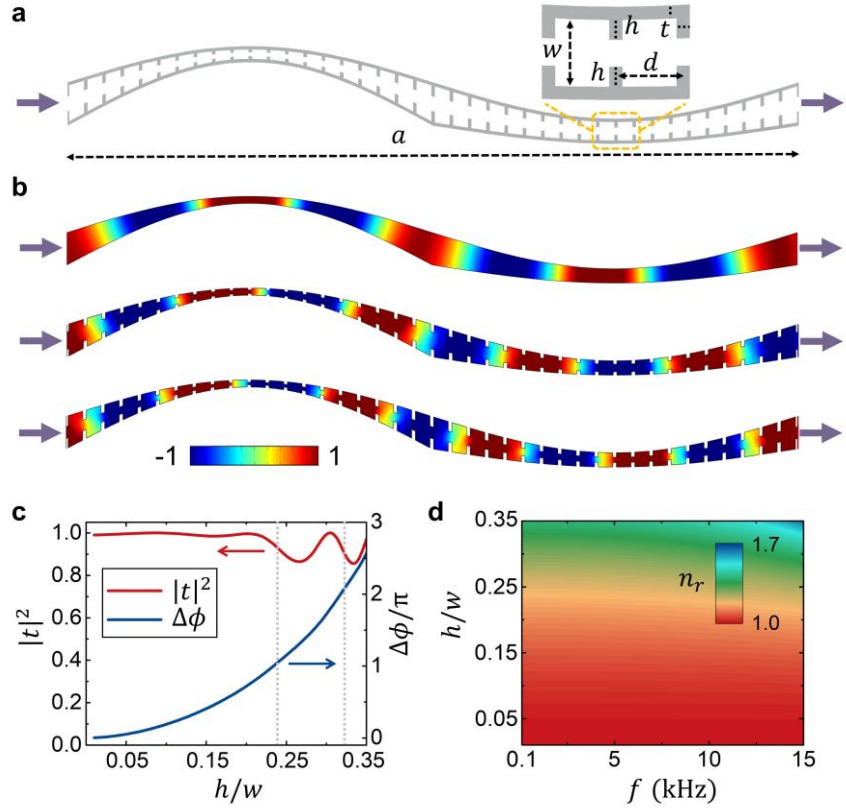

**Fig. 2. Design of acoustic tunnels with tunable acoustic paths. a** Schematic diagram of a tunnel with a varying cross-section along a curved path. The grey portion represents the solid structure. A magnified view is displayed in the top-right corner. Arrows indicate the incidence and transmission of waves. **b** Simulated acoustic pressure field distributions that vary with the ratio of $h/w$, corresponding to phase shift $\Delta\phi = 0, \pi, 2\pi$. **c** Phase shift $\Delta\phi$ and transmittance $|t|^2$ under normal incidence as a function of $h/w$. Vertical grey dotted lines indicate the cases with $\Delta\phi = \pi, 2\pi$. **d** Equivalent relative refractive index $n_r$ as a function of $h/w$ and the frequency $f$.

**Design of the ultra-broadband illusion metamaterial and experimental setup**

This illusion metamaterial as a SS device is constructed by assembling an array of such tunnels to route around the obstacles and guide sound waves from the entry surface to the exit other. To create the ultra-broadband SS effect, there are two criteria. The first criterion is that the entry and exit surfaces of the illusion metamaterial should have exactly the same shape (dotted lines in Fig. 1b). The other criterion is that all the tunnels should have exactly the same acoustic path. If the two criteria are met by the illusion metamaterial, the wavefront of the incident acoustic field can be copied from the entry to the exit surface, therefore realizing the two illusionary effects of "*disappearing space*" and "*time shift*".

Figure 3a shows the schematic diagram of a typical example of the illusion metamaterial and the corresponding experimental setup. There are three rhombic hard obstacles arranged along the *y* direction in air. The entry and exit surfaces of the illusion metamaterial are both flat surfaces in this case. A total number of twenty tunnels are applied for each rhombic obstacle, as shown by the zoom-in figure in Fig. 3b. Because of the mirror symmetry, ten different sets of $h/w$ are designed, as shown in Fig. 3c. Such parameters can guarantee that the acoustic path is almost the same ($L = 22.5$ cm) for all twenty tunnels, despite that they have obviously different lengths and bending angles. In experiment, the illusion metamaterial as well as the obstacles are both fabricated by using 3D printing techniques. A photo of the sample and the experimental setup is shown in Fig. 3d. The whole experiment is performed in a plate waveguide with a height of 3 cm. A cylindrical wave is emitted by a speaker located at a distance of $d_1 = 20$ cm away from the sample. A microphone is mounted on a horizontal translation stage to scan the transmitted signals located at the black dashed area shown in Fig. 3a in the $xy$ plane. The measured area with the size of $\Delta x = 10$ cm and $\Delta y = 20$ cm is of $d_2 = 5$ cm away from the sample. Sound absorbing foams are placed around the platform to reduce the reflected waves and noise from the environment.

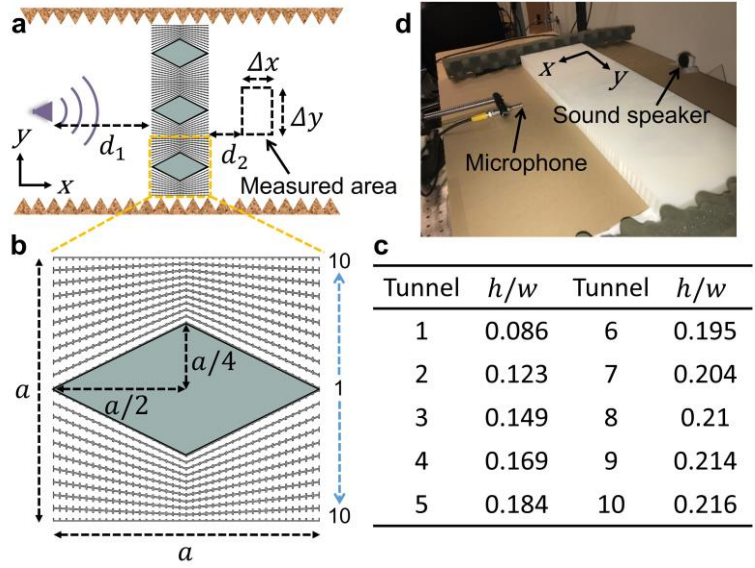

**Fig. 3. Design of the ultra-broadband illusion metamaterial and experimental setup. a** Schematic diagram of the illusion metamaterial and experimental setup. A magnified view of the metamaterial is shown in **b**. The side length of the square-shaped unit is $a = 20$ cm. The diagonal lengths of the rhombic-shaped obstacle are represented by $a$ and $a/2$, respectively. **c** Geometric parameter $h/w$ for various tunnels in the metamaterial. **d** Picture of the experimental setup.

**Ultra-broadband illusionary effect of "*disappearing space*" and sound scattering suppression**

In Fig. 4, we demonstrate the full-wave simulation and acoustic experimental results obtained by using the illusion metamaterial designed in Fig. 3. For the case of bare obstacles at $f = 4860$ Hz (Fig. 4a), large scattering clearly occurs due to the huge impedance mismatch between obstacles and air, as confirmed by the shadows behind the obstacles. However, when the obstacles are embedded in the designed SS device (Fig. 4b), the impinging cylindrical waves are guided around the obstacles, and then reproduce the cylindrical wavefront of the incident waves in the transmission region. From the wavefront in the transmission region, the point source seems to have moved forward from the real point of $(-30\text{ cm}, 0\text{ cm})$ to a virtual point of $(-10\text{ cm}, 0\text{ cm})$. Such a position change of 20 cm corresponds to the side length of the illusion metamaterial. Therefore, an illusion of "*disappearing space*", i.e. the whole space of the metamaterial with the embedded obstacles has completely disappeared, is proved. Moreover, the field on the incidence region is almost undisturbed, indicating that the reflection is negligibly small. The measured results, as shown in the right inset graphs of Figs. 4a and 4b, agree well with the full-wave simulations.

To demonstrate the ultra-broadband property, the numerical and experimental results obtained for two more frequencies, i.e. $f = 6860$ Hz and $f = 8860$ Hz are shown in Figs. 4c-4f. We can observe that the illusionary effects of "disappearing space" maintain perfectly under these frequencies with the illusion metamaterial, proving the ultra-broadband property.

To quantify the performance of illusion metamaterial in a broad spectrum, we calculate the normalized scattered pressure intensity, which is defined as $\gamma_{t,r} = |p_s|^2/|p_0|^2$, where $p_s, p_0$ are the scattered field and the empty field, respectively. The subscript $t$ and $r$ represent the transmission region ($x \in (10\text{ cm}, 50\text{ cm})$) and incidence region ($x \in (-50\text{ cm}, -10\text{ cm})$), respectively. The scattered field $p_s$ is obtained by subtracting the empty field without the sample ($p_0$) from the total field with the sample ($p$). It should be noted that, due to the illusionary effects, the empty field in the transmission region is calculated by using the virtual point source located at $(-10\text{ cm}, 0\text{ cm})$. On the other hand, the empty field for the incidence region is calculated by using the real point source located at (−30 cm, 0 cm). The results are plotted in Figs. 4g-4h. It is clearly seen that $\gamma_{t,r}$ maintain almost zero with the illusion metamaterial in an ultra-broad spectrum 1 kHz~16 kHz. In contrast, $\gamma_{t,r}$ are significantly larger in the case of bare obstacles.

We emphasize that the wide bandwidth of SS in this approach is far beyond any other previous approaches. Previous methods of realizing SS as "empty space" (*27–34*) usually have narrow bandwidths, except for ray optics cloaks (*44, 45*) that have ignored phase difference and thus cannot be described as a SS device. Here, the illusion metamaterial works for all frequencies below an upper limit where more than one propagating mode occur inside the tunnels. We note that this functionality of SS is independent of the wavefront of the incident waves. More examples such as the Gaussian wave incidence and multiple point source incidence are also plotted in the Supplemental Materials, which clearly verify the universality of this principle.

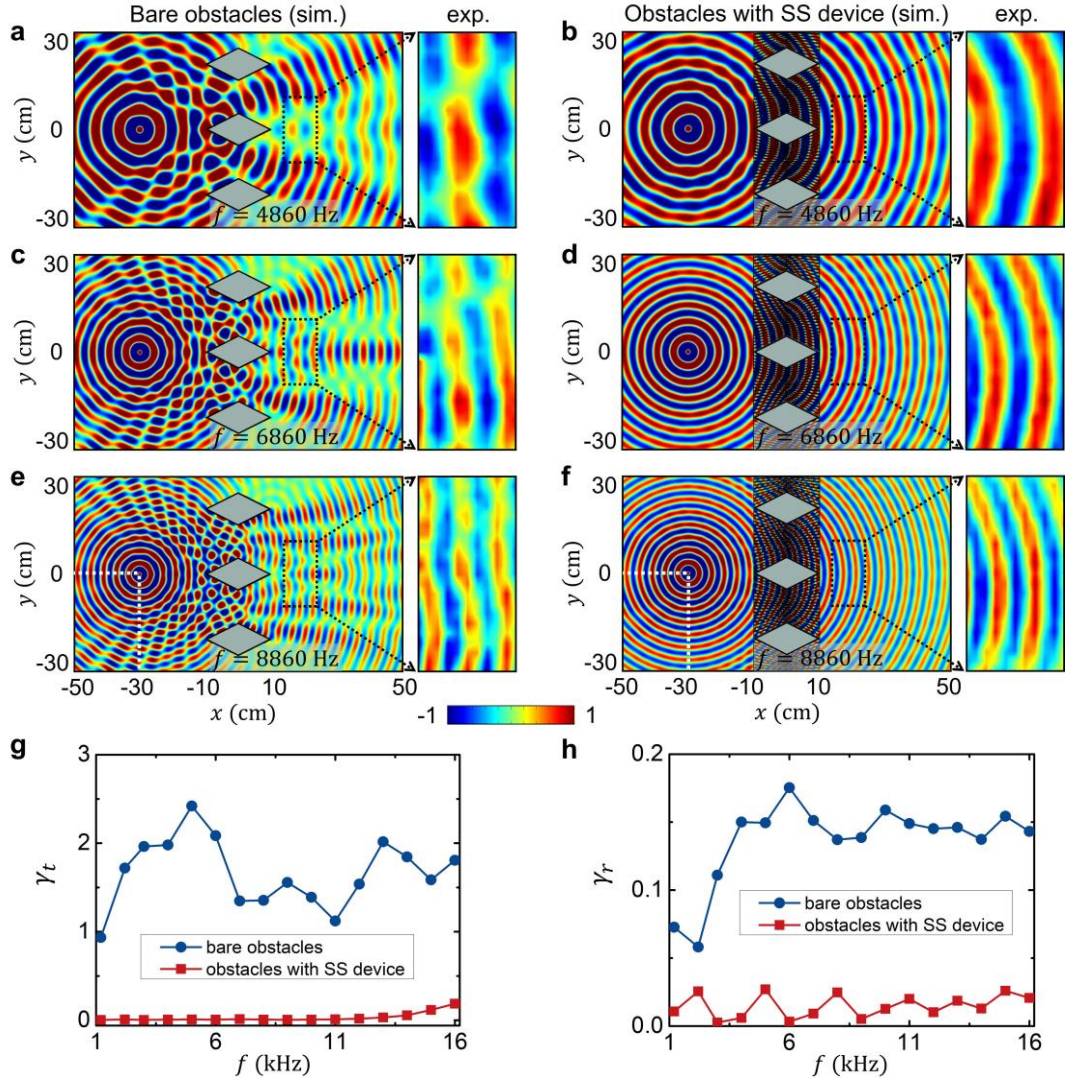

**Fig. 4. Ultra-broadband illusionary effect of "*disappearing space*" and sound scattering suppression. a,c,e** Simulated acoustic pressure field distributions for the scenario of bare obstacles under a point source radiation at different operating frequencies of $f = 4860$ Hz, 6860 Hz, 8860 Hz, and the corresponding experimental measurements (located in the black dashed area) are on the right. **b,d,f** Results for the scenario of obstacles covered by the SS device. **g-h** Frequency dependence of the normalized scattered pressure intensity for the transmission region ($\gamma_t$) and reflection region ($\gamma_r$).

**Illusionary effect of "*time shift*"**
We emphasize that besides the illusionary effect of "*disappearing space*", there is, simultaneously, another illusionary effect of "*time shift*". This illusionary effect can be demonstrated by considering a sound pulse propagating through a collection of obstacles with and without the illusion metamaterial, as shown in Fig. 5. A number of rhombic obstacles are arranged along the $y$ direction. The pulse is a time-domain Gaussian signal with a broad bandwidth. Clearly, the scattering by the obstacles has produced intense reflected waves and significantly changed the wavefront in transmission in the case of bare obstacles, as shown in Fig. 5a. While with the illusion metamaterial (Fig. 5b), the reflected waves are much smaller, because of the good impedance matching between the metamaterial and background medium (free space). In the transmission region, the circular wavefront is perfectly maintained, as if emitted from a shifted virtual point source located at $(-10\text{ cm}, 0\text{ cm})$. Furthermore, by comparing the wavefronts in the incidence and transmission regions, it is seen that there is a "*time shift*" in the transmission region. This is because that although the equal acoustic path in the tunnels leads to the illusionary effect of "disappearing space", but such a "disappeared space" still takes a finite time for the waves to pass. The value of the extra time shift is obtained as $t_{\text{shift}} = L/c_0 = 0.66$ ms, which is consistent with the results obtained from Fig. 5b. The zoom-in inset graphs on the right of Fig. 5b exhibit the details of how the wavefront of the sound waves are routed around the obstacles inside the metamaterial, thereby significantly reducing the scattering effect in Fig. 5a. More details are shown in Supplemental Materials. The length of time shift can be modulated by controlling the acoustic path. The corresponding videos are shown in Supplemental Videos.

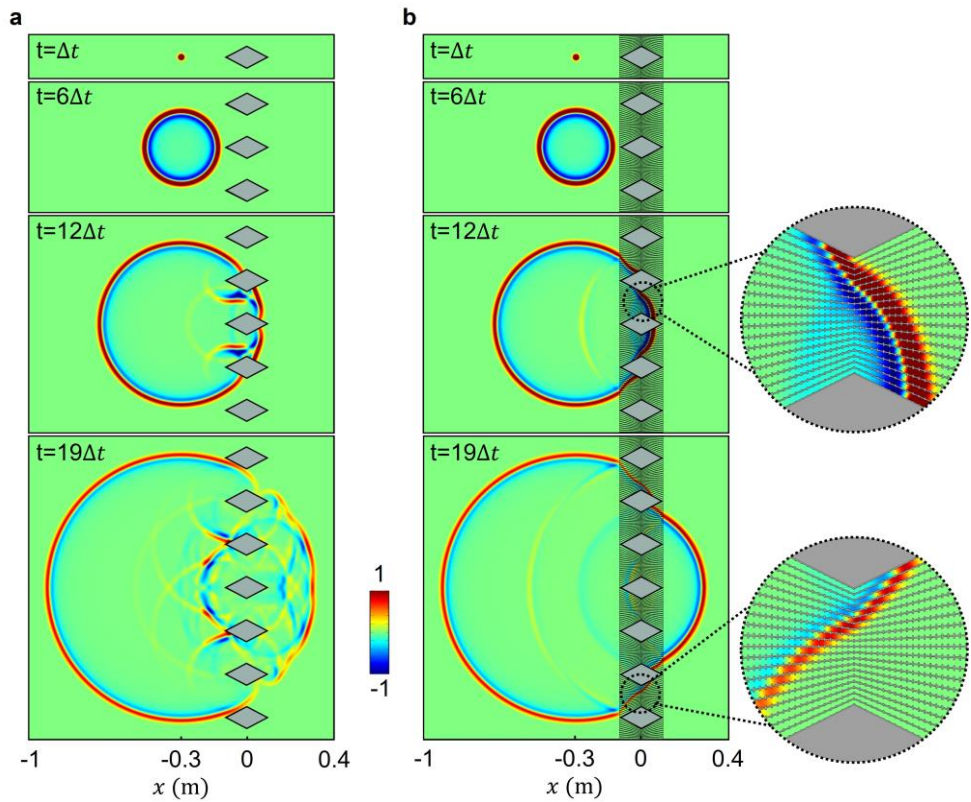

**Fig. 5. Pulse radiation and illusionary effect of "*time shift*".** **a** Snapshots of a sound pulse propagating through a collection of obstacles with significant scattering. **b** Snapshots of a sound pulse propagating through the obstacles embedded in the illusion metamaterial. The inset graphs display magnified details of the wavefront inside the metamaterial. The numbers indicate the order of the snapshots, where $\Delta t = 0.1$ ms. See Supplemental Videos.

**Illusion metamaterial and SS for a random collection of obstacles**
As a further demonstration of the robustness of this approach, we design a particular illusion metamaterial to suppress the sound scattering from three obstacles in a random arrangement. As shown in Fig. 6a, the illusion metamaterial is set in a square shape with a side length of $2.25a$ ($a = 20$ cm), and is composed of 46 tunnels, whose geometric parameters are presented in the Supplemental Materials. Here, the uniform acoustic path in these tunnels is set as $L = 47.5$ cm. The size of the rhombic obstacles is the same as above. From the field distributions plotted in Fig. 6b, it is clearly seen that when the obstacles are covered by the illusion metamaterial, the wavefronts in both the incidence and transmission regions become undistorted, as if the obstacles disappear. The illusion effects of "*disappearing space*" of length 45 cm and a "*time shift*" of 1.38 ms are both observed in the transmission region. On the contrary, mussy scattering occurs in the case of bare obstacles. Field distributions are shown in the Supplemental Materials. Furthermore, the normalized scattered pressure intensity is calculated and shown in Figs. 6c-6d, which turns out to be enormously reduced in the regime of 1 kHz-16 kHz.

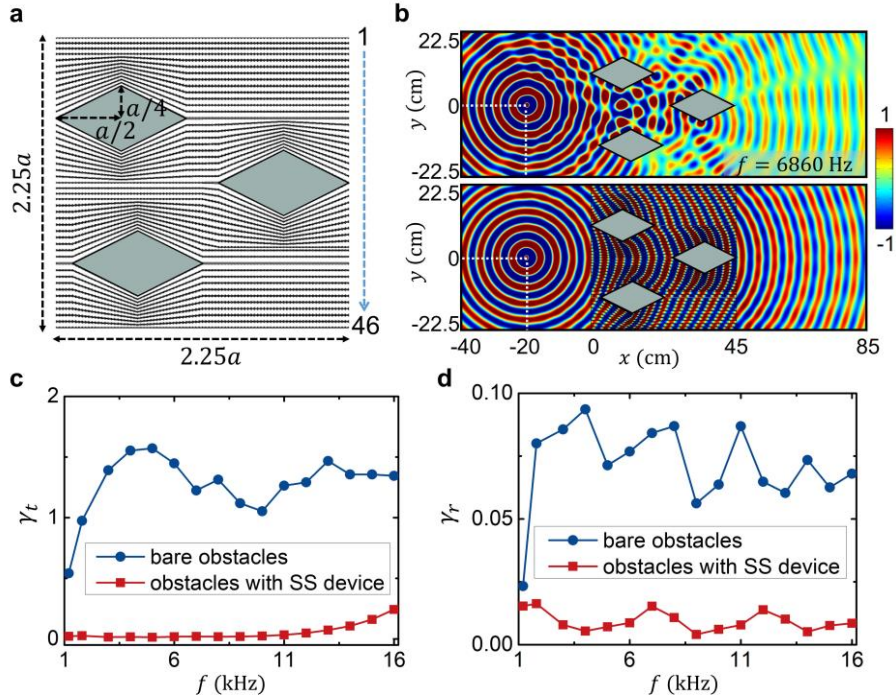

**Fig. 6. Ultra-broadband illusion metamaterial and sound scattering suppression for a random collection of obstacles. a** Schematic of a specific illusion metamaterial composed of three rhombic obstacles arranged randomly. **b** Simulated acoustic pressure field distributions for the scenarios of obstacles with or without the SS device under a point source radiation at an operating frequency of $f = 6860$ Hz. The point source is located at (-20 cm, 0 cm), and the center of three obstacles are located at (10 cm, 10 cm), (35 cm, 0 cm) and (12.5 cm, -12.5 cm), respectively. **c-d** Frequency dependence of the normalized scattered pressure intensity for the transmission region ($\gamma_t$) and reflection region ($\gamma_r$).

## DISCUSSION

It is recognized that when the tunnel width is small, e.g. <1mm, dissipation effect cannot be ignored due to the viscous friction between the air and the hard boundaries. On the other hand, the illusionary effects are impaired when the tunnels support more than one propagating mode, or when the wavelength is small enough to cause undesirable diffraction effects (*1*, *2*). Therefore, the operating bandwidth has an upper limit, which is around 16 kHz in our design. Interestingly, there is no lower limit for our design and the influence of loss can be reduced to a negligible level by applying relatively large tunnel widths as well as shorter lengths.

Causality is one of the essential reasons why previous illusions of "empty space" have a limited bandwidth (*44*). However, in this work, there is no similar constraint due to the existence of the "*time shift*", which, remarkably, enables ultra-broad operating bandwidth. It's worth mentioning that the illusion of "*disappearing space*" is also fundamentally different from that of "empty space", which was extensively studied in invisibility cloaks and other works. In other words, we are not realizing an invisibility cloak here. Nevertheless, all the consequences of wave scattering are completely removed, accompanied by the ultra-broadband illusionary effects of "*disappearing space*" and "*time shift*". It is therefore an ideal SS device.

The most striking feature of such metamaterial is its almost frequency-independent functionality over an ultra-wide range of frequencies from the quasi-static limit to an upper limit. This key feature may find wide applications in vital applications such as stealth and reverberation control. Our work thus enables a powerful platform for the long-desired broadband sound manipulation.

## MATERIALS AND METHODS

**Numerical simulations**

The finite element software COMSOL Multiphysics is performed for the full-wave simulations with "Pressure Acoustics, Frequency Domain" and "Pressure Acoustics,

Transient" modules. The mass density and sound velocity of air are set as $1.21 \text{ kg/m}^3$ and $343 \text{ m/s}$, respectively. The resin structures are treated as acoustically rigid materials. The plane wave radiation is used in Fig. 2, and the monopole point source is used in Figs. 4-6. Perfectly matched layers are adopted to reduce the reflections.

**Experimental measurements**

All samples are fabricated with resin by using stereolithography 3D printing techniques (SLA, 0.1 mm in precision). A sound speaker radiates eight periods of sound waves with single frequency we want. A microphone (PCB 130F20) scans the sound field located at the black dashed area shown in Fig. 3a. The scanning has resolution of 1 cm and in total 231 points are scanned. The measurement platform is a 2D waveguide with a height of 30 mm. The sound absorbing foams are set around the system to reduce the reflections.

**SUPPLEMENTAL MATERIALS**

Supplemental material for this article is available at …

**Acknowledgments:** C.L. thanks J. Luo for helpful discussions. **Funding:** Y.L. acknowledges the support of the National Key Research and Development Program of China (Grant No. 2020YFA0211400) and the National Natural Science Foundation of China (Grants No. 11974176) for this work. **Author contributions:** C.L. and C.M. contributed equally to this work. Y.L. and N.X.F. supervised the project. C.L. and Y.L. conducted the analysis, simulations and sample fabrication. C.M. and N.X.F. conducted the experiment design and measurements. All the authors contributed to the data analysis and manuscript preparation. **Competing interests:** The authors declare that they have no competing interests. **Data and materials availability:** All data needed to evaluate the conclusions in the paper are present in the paper and/or the Supplementary Materials.